\documentclass[
    ,final            % use final for the camera ready runs
  ]
  {aipproc}

\layoutstyle{8x11single}

\begin{document}

\title{News on  Penguins\footnote{Based on an invited talk given at the 19th Particles and Nuclei International Conference
PANIC11, MIT, Cambridge, USA, 24th-29th July 2011}}

\classification{13.20.He, 12.38.Bx, 12.39.St. \hspace{8.5cm} MZ-TH/11-32}

%\keywords      {Rare B decays, QCD corrections, Indirect search for new physics.}
\author{Tobias Hurth}{
  address={Institute for Physics (THEP),  Johannes Gutenberg University, D-55099 Mainz, Germany}  
}

\begin{abstract}
 We summarize recent theoretical developments in the field of radiative and semileptonic
 penguin decays.
\end{abstract}

\maketitle

%%%%%%%%%%%%%%%%%%%%%%%%%%%%%%%%%%%%%%%%%%%%
%% MAINMATTER
%%%%%%%%%%%%%%%%%%%%%%%%%%%%%%%%%%%%%%%%%%%%

\section{Ambiguity of the New Physics Scale}
Within the indirect search for New Physics (NP)  there is a an ambiguity of the new physics  scale. In the model-independent approach using the effective electroweak 
hamiltonean, the contribution to one specific operator ${\cal O}_i$ can be parametrized via
\begin{equation}
( C^i_{\rm SM}\, / \,  M_W  +  C^i_{\rm NP}\, / \, \Lambda_{\rm NP} ) \times {\cal O}_i 
\end{equation}
where the first term represents the SM contribution at the electroweak scale $M_W$ and the second one the NP contribution with an unknown coupling $C^i_{\rm SM}$ and an unknown 
NP scale $\Lambda_{\rm NP}$.

The radiative and semi-leptonic penguin modes,  $b \rightarrow s \gamma$  and $b \rightarrow  s  \ell^+\ell^-$, 
are  flavour changing neutral current (FCNC) processes and, thus,   are highly sensitive for new degrees of freedom via virtual effects (for  reviews, see
\cite{Hurth:2010tk,Hurth:2007xa,Hurth:2003vb})
The non-existence of large NP effects in flavour observables in general~ \cite{Buchalla:2008jp,Antonelli:2009ws} 
 implies the
famous flavour problem, namely why FCNC are suppressed. 
Either the mass scale of the new degrees of freedom, $\Lambda_{\rm NP}$,  is very high or the 
new flavour-violating couplings, $C^i_{\rm NP}$,  are small for (symmetry?) reasons that remain to be found. 
For example,  assuming  {\it generic\/}  new flavour-violating couplings of $O(1)$,  
the present data on  $K$-$\bar K$ mixing implies  a very high NP scale of order $10^3$--$10^4$ TeV 
depending on whether the new  contributions enter at loop- or at tree-level.   
 In contrast, theoretical  considerations on the Higgs sector, which is responsible for the mass generation 
of the fundamental particles in the SM, call for NP at
order $1$ TeV.    As a consequence,
any NP below the  $1$ TeV scale must have a non-generic flavour structure.  
In addition, also the present electroweak data indicate a slightly higher NP scale, a data-driven problem 
known as little hierarchy problem. 

The present measurements of $B$ decays, especially of FCNC
processes, already significantly restrict the parameter
space of NP models. In general such bounds from flavour physics are model-dependent, but often 
much stronger than the ones derived from other measurements. In any case, the indirect flavour
information will be most valuable when the general nature and the mass scale of NP will be known.

\section{The inclusive decay $\bar B \rightarrow X_{s,d}  \gamma$}

{\it Perturbative contributions:}  Among the rare decay modes, the inclusive decay $\bar B \rightarrow X_s \gamma$  is the 
most important one, because it is theoretically 
well-understood and at the same time it  has been
 measured extensively at the $B$ factories.
While non-perturbative  corrections to this decay mode are subleading and recently estimated to be well below $10\%$~\cite{Benzke:2010js}, 
perturbative QCD corrections are the most important corrections.
Within a global effort,  a  perturbative QCD calculation to the next-to-next-to-leading-logarithmic 
order level (NNLL) has quite recently been
performed and has led to the first NNLL prediction of the $\bar B \to X_s  \gamma$ branching 
fraction~\cite{Misiak:2006zs} with a photon cut at $E_\gamma = 1.6 {\rm GeV}$ (including the error due to nonperturbative corrections):
\begin{equation}\label{final1}
{\cal B}(\bar B \to X_s \gamma)_{\rm NNLL} =  (3.15 \pm 0.23) \times 10^{-4}.
\end{equation}
This result is based on various highly-nontrivial perturbative
calculations ~\cite{Misiak:2004ew,Bobeth:1999mk,Gorbahn:2004my,Gorbahn:2005sa,Czakon:2006ss,Blokland:2005uk,Melnikov:2005bx,Asatrian:2006ph,Asatrian:2006sm,Bieri:2003ue,Misiak:2006ab}. 
The combined experimental data leads to 
(Heavy Flavor  Averaging Group (HFAG)~\cite{hfag}) 
\begin{equation}
 {\cal B}(\bar B \rightarrow X_s \gamma) = (3.55
  \pm 0.24 \pm 0.09) \times 10^{-4}, 
\end{equation}
where the first error is combined statistical and systematic, and the second
is due to the extrapolation in the photon energy.
Thus, the SM prediction and the experimental average are consistent at the $1.2 \sigma$ level.
This is one important  example that the CKM theory is not only confirmed by 
the data entering into the CKM unitarity fit, but also by many additional flavour mixing phenomena.

{\it Nonperturbative contributions:}  It was noted long ago~\cite{Ligeti:1997tc}, that there is no local OPE for the
inclusive decay $\bar B \rightarrow X_s \gamma$ if one considers
operators beyond the leading electromagnetic dipole operator 
${\cal O}_7$.  Then, there are  so-called  {\it resolved}\/ photon contributions which contain 
subprocesses in which the photon couples to light partons
instead of connecting directly to the effective weak-interaction vertex~\cite{Benzke:2010js}.
Only recently,  a  systematic analysis~\cite{Benzke:2010js} of
all resolved photon contributions related to other operators in the weak
Hamiltonian has  established this breakdown of the local OPE within the
hadronic power corrections as a generic result.
Clearly, estimating such nonlocal matrix elements is very difficult, and
an irreducible theoretical uncertainty of $\pm
(4-5)\%$ for the total $CP$ averaged decay rate, defined with a
photon-energy cut of
$E_\gamma = 1.6$ GeV, remains~\cite{Benzke:2010js}. This result strongly
indicates that the theoretical efforts for the $B \to X_s \gamma$
mode have reached the nonperturbative boundaries, but it also reconfirms the 
dominance of the perturbative contributions. 

There is another positive result induced by this new analysis:
Until recently, it was believed that the long-distance contributions from the intermediate $u$-quark in
the penguin loops are critical.  
They are suppressed in the $\bar  B  \rightarrow X_s \gamma$ mode by the CKM matrix elements. In
$\bar  B \rightarrow X_d \gamma$, there is no CKM suppression, and one must account for 
the nonperturbative contributions that arise from the $u$-quark loops. A simple dimensional estimate 
leads to an uncertainty of at least  $10\%$.  
However,  this  contribution  vanishes in the total
$CP$-averaged  rate of $\bar B \to X_s \gamma$ at order $\Lambda/m_b$~\cite{Benzke:2010js}. This result applies to 
the total rate of $\bar B \to X_d \gamma$ as well. Thus, there is no  power correction due to the $u$-quark loops 
in the total rate of $\bar B \to X_d \gamma$ at order $\Lambda/m_b$, which implies that the $CP$-averaged 
decay
rate of $\bar B \rightarrow X_d \gamma$ is as theoretically clean as the decay rate 
of $\bar B \rightarrow X_s \gamma$~\cite{Hurth:2010tk}.
The present NLL-prediction reads~\cite{Hurth:2003dk,Hurth:2003pn}  (for $E_\gamma > 1.6 \, {\rm GeV}$): 
\begin{equation} 
{\cal B} (\bar B \to X_d \gamma)
=  \left( 1.38 \,\,  \left. {}^{+0.14}_{-0.21}   \right|_{m_c \over m_b}
                     \pm 0.15_{\rm CKM} \pm 0.09_{\rm param.} \pm 0.05_{\rm scale} \right) \times 10^{-5}  \,.
\end{equation}
The large CKM-uncertainty is due to $V_{td}$. The uncertainty due to the charm-scheme dependence can 
be reduced significantly by a NNLL analysis in analogy to $\bar B \rightarrow X_s \gamma$. 
Already the NLL prediction in combination with the first real  measurement of this inclusive mode~\cite{:2010ps}
leads to interesting bounds  on NP~\cite{Crivellin:2011ba}.

But  there is also a negative consequence: Without CP-averaging the non-perturbative the $u$-quark contribution 
survives and dominates the direct CP asymmetry in $\bar B \rightarrow X_{s,d} \gamma$. Perturbative contributions
exhibit a triple suppression; one has an $\alpha_s$ suppression in order  to have a strong phase, a Cabibbo suppression of $\lambda^2$,
and  a GIM suppression of $(m_c/m_b)^2 \approx \Lambda/m_b$ reflecting the fact that in the limit $m_c \rightarrow 0$ 
the direct CP violation vanishes.  Thus, the non-perturbative (resolved) $u$-quark contribution is enhanced by a factor $1/\alpha_s$
compared to the perturbative contributions. Model estimates of the resolved contribution lead to~\cite{Benzke:2010tq}
\begin{equation}
-0.6\%<{\cal A}(\bar B \rightarrow X_s\gamma)^{\rm SM}<2.8\%, 
\end{equation}
which covers most of the experimentally allowed range~\cite{hfag}:  ${\cal A}(\bar B \rightarrow X_s\gamma)   = - (1.2\pm 2.8)\% \,.$
But the untagged direct CP asymmetry {\it survives}. All resolved contributions cancel at order $\Lambda/m_b$, so we still have 
the zero-prediction for the untagged CP asymmetry for $\bar B \rightarrow X_{s+d} \gamma$~\cite{Soares:1991te,Hurth:2001yb,Hurth:2003dk}.
This prediction is directly based on the CKM-unitarity and on the smallness of the U-spin breaking parameters $(m_s/m_b)^2$. 
Thus, it represents a clean test whether  new CP phases are active or not.

{\it There are still some open issues  in the decay $\bar B \rightarrow X_s \gamma$}  which ask for further study. 
First,   there are  three-loop matrix elements of the
leading four-quark operators, which have first been calculated within the
so-called large-$\beta_0$ approximation~\cite{Bieri:2003ue,Ligeti:1999ea,Boughezal:2007ny}.  A
calculation that goes beyond this approximation by employing an
interpolation in the charm quark mass $m_c$ from $m_c > m_b$ to the
\mbox{physical} $m_c$ value has been presented in
Refs.~\cite{Misiak:2006ab,Misiak:2010sk}. In this interpolation the $\alpha_s^2
\beta_0$ result~\cite{Bieri:2003ue} is assumed to be a good
approximation for the complete $\alpha_s^2$ result for vanishing charm
mass. It is this part of the NNLL calculation which is still open for
improvement.  Indeed, there are several collaborations presently working on this issue. 
A complete calculation of the three-loop matrix
elements of the four-quark operators ${\cal O}_{1,2}$ for vanishing
charm mass is in progress  and will
cross-check the  error estimate due to
the interpolation~\cite{BoughezalII,CzakonI}.  Partial results are already available~\cite{Schutzmeier}.  
In addition, there is an  effort~\cite{CzakonII} to calculate the matrix elements 
for arbitrary $m_c$ directly.  All these efforts will finally eliminate the $3\%$ uncertainty due to the 
interpolation within the present NNLL prediction~\cite{Misiak:2006zs}.

Second, in the measurement of the inclusive mode $\bar B \to X_s \gamma$ one
needs cuts in the photon energy spectrum to suppress the background from
other $B$ decays which  induces additional sensitivities to non-perturbative physics. 
These shape-function effects were  taken into account in the experimental
analysis. The corresponding theoretical uncertainties due to this
model dependence are  reflected in the extrapolation error of  the
experimental results. The extrapolation is done 
from the experimental energy cut values down 
 to $1.6\;{\rm GeV}$.
But a cut around $1.6\,{\rm GeV}$ might not guarantee that a
theoretical description in terms of a local OPE is sufficient because of
the sensitivity to the additional scale $\Delta = m_b - 2 E_\gamma$~\cite{Neubert:2004dd}. 
A multiscale OPE with three
short-distance scales $m_b, \sqrt{m_b \Delta}$, and $\Delta$  has been 
proposed to connect the shape function and the local OPE region.
Recently, such additional perturbative cutoff-related effects have been 
calculated to NNLL precision by the use of SCET
methods~\cite{Becher:2006pu,Becher:2005pd,Becher:2006qw}.  Such
perturbative effects due to the additional scale are negligible at 
$1.0\,{\rm GeV}$ but of order $3\%$ at $1.6\,{\rm  GeV}$~\cite{Becher:2006pu}.  
The size of these effects at $1.6\,{\rm
  GeV}$ is similar to the $3\%$ higher-order uncertainty in
the present NNLL prediction.  However, the numerical consistency of the
SCET analysis  has  recently been questioned~\cite{Misiak:2008ss}.  
Far away from the endpoint ($E_0=1.6$ GeV),
the logarithmic and nonlogarithmic terms cancel; this kind of cancellation was already  observed  
 in Ref.~\cite{Andersen:2006hr}. Within the resummation of the
cutoff-enhanced logarithms this feature leads to an overestimate of the
$O(\alpha_s^3)$ terms~\cite{Misiak:2008ss}.  Further
work is needed to clarify this issue.

Third, the $\bar B \rightarrow X_s \gamma$ decay rate is  normalized to the {\it charmless} semileptonic
rate in order to separate the charm  dependence. Then the quantity  
$C = |V_{ub}|^2 / |V_{cb}|^2 \,\times\, \Gamma[B\to X_c e  \bar \nu] / \Gamma [ B\to X_u e \bar\nu] $  enters the theoretical prediction. 
  Recently, a  scheme dependence in the
determination of the prefactor $C$ was noticed~\cite{Gambino:2008fj},  which is around $3\%$, thus
within the perturbative uncertainty~\cite{Misiak:2008ss}. The two determinations in the 
$1S$ scheme~\cite{Bauer:2004ve} and in the kinetic scheme~\cite{Gambino:2008fj} differ through renormalization 
schemes, methodology, and experimental input. An update of the analysis 
within the $1S$ scheme might resolve part of the corresponding uncertainty. 

\section{The inclusive decay $B \rightarrow X_s \ell^+\ell^-$}
{\it Perturbative contributions:}  This inclusive mode is also dominated by perturbative contributions, if
 one eliminates $c\bar c$ resonances with the help of kinematic cuts.
In the so-called `perturbative $q^2$-windows' below and above the
resonances, namely in the low-dilepton-mass region $1\;{\rm GeV}^2 < q^2
= m_{\ell\ell}^2 < 6\;{\rm GeV}^2$, and also in the high-dilepton-mass
region with $q^2 > 14.4\;{\rm GeV}^2$, theoretical predictions for the
invariant mass spectrum are dominated by the perturbative contributions.
QCD corrections are calculated to NNLL 
precision~\cite{Asatryan:2001zw,Asatryan:2002iy,Ghinculov:2002pe,Asatrian:2002va,Gambino:2003zm,Ghinculov:2003bx,Bobeth:2003at,Ghinculov:2003qd,Greub:2008cy,
Huber:2007vv}.
More recently  electromagnetic corrections were   calculated: NLL quantum
electrodynamics (QED)  two-loop corrections to the
Wilson coefficients are of $O(2\%)$~\cite{Bobeth:2003at}. 
Also, in the QED one-loop corrections to matrix elements, large
collinear logarithms of the form $\log(m_b^2/m^2_\ell)$ survive
integration if only a restricted part of the dilepton mass spectrum is
considered. These collinear logarithms add another contribution of order $+2\%$ 
in the low-$q^2$ region for $B \to X_s \mu^+\mu^-$~\cite{Huber:2005ig}. For  the high-$q^2$ region,  one finds 
$-8\%$~\cite{Huber:2007vv}.

{\it Quark-Hadron-Duality:}  The integrated branching fraction of $\bar B \to X_s \ell^+\ell^-$  is dominated
by the  resonance background which exceeds the nonresonant charm-loop
contribution by two orders of magnitude.  
As has been recently noticed~\cite{Beneke:2009az},
this feature should not be
misinterpreted as a striking failure of global parton-hadron duality
which postulates that the sum over the hadronic final states, including
resonances, should be well approximated by a quark-level
calculation~\cite{Poggio:1975af}.  Crucially, the
charm-resonance contributions to the decay $\bar B \to X_s \ell^+\ell^-$
are expressed in terms of a phase-space integral over the absolute
square of a correlator.  For such a quantity global quark-hadron
duality is not expected to hold.  Nevertheless,
local quark-hadron duality (which, of course, also implies global duality) 
  may be reestablished by resumming Coulomb-like
interactions~\cite{Beneke:2009az}.

\section{The exclusive decay  $B \to K^{(*)}\ell^+ \ell^-$}
{\it QCD factorization:} The method of QCD factorization  (QCDF)~\cite{Beneke:1999br}
 and its field-theoretical formulation SCET~\cite{Bauer:2000ew,Bauer:2000yr}
 form  the basis of the up-to-date predictions of exclusive
$B$ decays.  
There is also a factorization formula for the exclusive semileptonic $B$
decays, such  as  $B \to K^* \ell^+ \ell^-$~\cite{Beneke:2001at,Beneke:2004dp}. 
The hadronic form factors
can be expanded in the small ratios $\Lambda/m_b$ and $\Lambda/E$, where
$E$ is the energy of the light meson.  If we neglect  corrections of order
$1/m_b$ and $\alpha_s$, the {\it seven}\/ a priori independent $B\to K^*$
form factors reduce to {\it two}\/ universal form factors $\xi_{\bot}$ and
$\xi_{\|}$~\cite{Charles:1998dr}.  
%This reduction makes it possible
%to design interesting ratios of observables in which any soft form
%factor dependence cancels out for all dilepton masses $q^2$ at leading
%order in $\alpha_s$ and $\Lambda/m_b$~\cite{Egede:2008uy}. However,
%the unknown $\Lambda/m_b$ corrections  restrict the NP sensitivity of such 
%observables.  
These  theoretical simplifications of the QCDF/SCET approach are valid 
to the kinematic region in which the energy of the $K^*$ is of the order
of the heavy quark mass, that is, $q^2 \ll m_B^2$. Thus, 
factorization formula applies well in the dilepton mass
range $1\; {\rm GeV}^2 < q^2 < 6\; {\rm GeV}^2 $.

{\it Full angular analysis:}  In the near future, a full angular
 analysis will become  possible. This  rich information
 allows for the design of observables with specific NP sensitivity and
reduced hadronic uncertainties~\cite{Egede:2008uy,Egede:2010zc}.  These
observables are 
constructed in such a way that the soft form factor dependence cancels
out at leading order in $\alpha_s$ and in $\Lambda/m_b$ for all low dilepton masses, and they 
have much higher sensitivity to new right-handed
currents than observables that  are already accessible via the
projection fits~\cite{Kruger:2005ep,Egede:2008uy,Egede:2010zc}.
In these optimized
observables, the unknown $\Lambda/m_b$ corrections are  the source of the
largest uncertainty.  Further detailed NP analyses of such  angular observables 
 have been   presented in Refs.~\cite{Bobeth:2008ij,Altmannshofer:2008dz}.  
A full angular analysis provides high sensitivity to various Wilson coefficients, but 
the sensitivity to new weak  phases is restricted, mainly due to large experimental 
uncertainties~\cite{Egede:2009tp,Egede:2010zc}.  
Observables defined at  high-$q^2$  also represent very interesting observables due to the fact
that $\Lambda/m_b$ corrections can be estimated with the help of heavy-quark effective theory (HQET), but
formfactors at high-$q^2$ have to be presently extrapolated from the low-$q^2$ region~\cite{Grinstein:2004vb,Beylich:2011aq,Bobeth:2010wg}.

%\begin{figure}
 % \includegraphics[height=.3\textheight]{golfer}
  %\caption{Picture to fixed height}
%\end{figure}

%\begin{table}
%\begin{tabular}{lrrrr}
%\hline
 % & \tablehead{1}{r}{b}{Single\\outlet}
 % & \tablehead{1}{r}{b}{Small\tablenote{2-9 retail outlets}\\multiple}
 % & \tablehead{1}{r}{b}{Large\\multiple}
%  & \tablehead{1}{r}{b}{Total}   \\
%\hline
%1982 & 98 & 129 & 620    & 847\\
%1987 & 138 & 176 & 1000  & 1314\\
%1991 & 173 & 248 & 1230  & 1651\\
%1998\tablenote{predicted} & 200 & 300 & 1500  & 2000\\
%\hline
%\end{tabular}
%\caption{Average turnover per shop: by type
 % of retail organisation}
%\label{tab:a}
%\end{table}

%%%%%%%%%%%%%%%%%%%%%%%%%%%%%%%%%%%%%%%%%%%%%%%%
%% BACKMATTER
%%%%%%%%%%%%%%%%%%%%%%%%%%%%%%%%%%%%%%%%%%%%%%%%

\begin{theacknowledgments}
TH thanks the organizers of the conference  for the interesting and 
valuable meeting  and the CERN theory group for its  hospitality during his regular visits to CERN where part of this work  was written. TH is also thankful
to Leonardo Vernazza for a careful reading of the manuscript.
\end{theacknowledgments}

\bibliographystyle{aipproc}

\end{document}